\documentclass[reqno,12pt]{article}
\usepackage{amsfonts,amssymb,amsmath,amscd,bm,bbold,cite,delarray,subfigure}
\usepackage{xypic}
\usepackage[dvips]{color}
\usepackage[dvips]{graphicx}

\numberwithin{equation}{section}

\font\capital=rsfs12
\font\scriptcapital=rsfs10 at 7 truept
\font\scriptscriptcapital=rsfs10 at 5 truept
\def\scri{\fam=15}
\newcommand{\mathscri}[1]{{{\scri #1}}}

\font\sansserif=cmss12
\font\scriptsansserif=cmss12 at 7 truept
\font\scriptscriptsansserif=cmss10 at 5 truept
\font\euler=eusm10 at 12 truept
\font\scripteuler=eusm7
\font\scriptscripteuler=eusm5 
\newcommand{\Beta}{{\mathrm{B}}}

\begin{document}

\hrule\vskip.4cm
\hbox to 14.5 truecm{June 2007\hfil DFUB 07}
\hbox to 14.5 truecm{Version 1  \hfil } %hep-th/yymmnnn}
\vskip.4cm\hrule
\vskip.7cm
\begin{large}
\centerline{\textcolor{blue}{\bf The Hitchin Model, Poisson-quasi-Nijenhuis} }  
\centerline{\textcolor{blue}{\bf Geometry and Symmetry Reduction}}  
\end{large}
\vskip.2cm
\centerline{by}
\vskip.2cm
\centerline{\textcolor{blue}{\bf\bf Roberto Zucchini}}
\centerline{\it Dipartimento di Fisica, Universit\`a degli Studi di Bologna}
\centerline{\it V. Irnerio 46, I-40126 Bologna, Italy}
\centerline{\it I.N.F.N., sezione di Bologna, Italy}
\centerline{\it E--mail: zucchinir@bo.infn.it}
\vskip.7cm
\hrule
\vskip.7cm
\centerline{\textcolor{blue}{  \bf Abstract} }
\par\noindent
\vskip.4cm
We revisit our earlier work on the AKSZ formulation of 
topological sigma model on generalized complex manifolds, or Hitchin model,  
\cite{Zucchini1}. We show that the target space geometry geometry implied by
the BV master equations is 
Poisson--quasi--Nijenhuis geometry 
recently introduced and studied by Sti\'enon and Xu (in the untwisted case) in
\cite{Xu2}. Poisson--quasi--Nijenhuis geometry is more general than generalized 
complex geometry and comprises it as a particular case. Next, we show how 
gauging and reduction can be implemented in the Hitchin model. We find 
that the geometry resulting form the BV master equation is closely related to but 
more general than that recently described by Lin and Tolman in \cite{Tolman1,Tolman2}, 
suggesting a natural framework for the study of reduction of 
Poisson--quasi--Nijenhuis manifolds. 

\vfill\eject

\begin{small}
\section{  \bf Introduction}
\label{sec:intro}
\end{small}
%\vskip.3cm
In type II superstring theory, an effective four dimensional 
low energy field theory is obtained by compactification of the 
six extra dimensions. In the absence of fluxes, requiring unbroken 
four dimensional $\mathcal{N}=2$ supersymmetry leads to the well known condition
that the six dimensional internal manifold should be Calabi--Yau.
In recent years, a large body of literature has been devoted to 
attempts to find a similarly elegant condition in the presence of NS and RR
fluxes, both for $\mathcal{N}=2$ and $\mathcal{N}=1$ supersymmetry. 
(See for instance \cite{Grana0} for a comprehensive review and 
extensive referencing). 
The intense scrutiny, which these more general compactifications have undergone,
reflects both their physical and mathematical interest. 

In flux compactifications of type II superstring theories, 
requiring unbroken four dimensional $\mathcal{N}=1$ supersymmetry 
leads to certain topological and differential conditions on the 
internal manifold $M$ \cite{Grana1,Grana2,Martucci1}. 
These conditions are naturally expressed in the mathematical language
of generalized complex geometry \cite{Hitchin1,Gualtieri}.
(See \cite{Zabzine3, Cavalcanti,Guttenberg} for recent reviews of this subject
aimed to a physical readership). They state the existence of two nowhere 
vanishing globally defined $TM\oplus T^*M$ pure spinors. One of these
satisfies the appropriate differential condition required for it to
define a twisted generalized Calabi--Yau structure on $M$.
The other, conversely does not, the obstruction being due to the 
presence of RR fluxes and warping.

Ordinary fluxless type II compactifications are described by  
$(2,2)$ superconformal sigma models on Calabi--Yau manifolds. 
These are however nonlinear interacting field theories and, so, are rather 
complicated and difficult to study. In 1988, Witten showed that a $(2,2)$ 
supersymmetric sigma model on a Calabi--Yau manifold could be twisted in 
two different ways, to yield the so called $A$ and $B$ topological sigma
models \cite{Witten1,Witten2}. Unlike the original untwisted sigma models, 
the topological models are soluble field theories: the calculation of 
observables can be carried out by standard methods of geometry and topology. 

The recent interest in flux compactifications has prompted the search for
topological sigma models on generalized complex manifolds. In the particular 
case of biHermitian manifolds \cite{Gates1}, this problem was tackled
in \cite{Kapustin1,Kapustin2} by Kapustin and Kapustin and Li, who 
formulated it in the suitable geometrical framework of generalized
Kaehler geometry \cite{Gualtieri} and derived the appropriate twisting 
prescriptions. In refs. \cite{Zucchini4,Zucchini5,Chuang}, 
developing on Kapustin's and Li's results, the biHermitian topological action
and symmetry variations were explicitly derived and written down. 

BiHermitian geometry can accommodate only NS flux. If one wishes to incorporate
RR fluxes, it is non longer sufficient. In the last few years, many attempts
have been made to construct topological sigma models with generalized complex 
target manifolds more general than generalized Kaehler ones 
\cite{Lindstrom2,Lindstrom3,Zucchini1,Zucchini2,Pestun1}.
All these endeavors were somehow unsatisfactory either because they
remained confined to the analysis of geometrical aspects of the sigma
models or because they yielded field theories which 
were not directly suitable for quantization.
In \cite{Zucchini1,Zucchini2,Pestun1}, the sigma models were constructed by 
employing the Batalin--Vilkovisky (BV) quantization algorithm \cite{BV1,BV2} in  
the Alexandrov--Kontsevich--Schwartz--Zaboronsky (AKSZ) formulation \cite{AKSZ}. 
To date, this seems to be the most promising approach to the solution of 
the problem of constructing interesting sigma models on generalized complex 
target manifolds, though, as shown in \cite{Zucchini3}, the implementation 
of gauge fixing remains a major technical obstacle even in the simplest cases.  

One efficient way of generating sigma models on non trivial manifolds 
is the gauging of sigma models on simpler manifolds. The target space 
of the gauged model turns out to be the quotient of that of the ungauged 
model by an action of the gauge group. In certain cases, when a 
symplectic structure and a moment map for the gauge group action
can be defined, this construction is a particular
case of a general procedure called Hamiltonian reduction \cite{Marsden}.
The gauging of (2,2) supersymmetric sigma models on biHermitian manifolds 
was studied originally by Hull, Papadopoulos and Spence in \cite{Spence1}
developing on the results of \cite{Gates1}. 
Their analysis was however limited to the subclass of almost product
structure biHermitian spaces because of the lack of an off--shell (2,2) 
supersymmetric action in the general case at that time. Recently, 
such action has been obtained in ref. \cite{Lindstrom5}. 
This has led the authors of \cite{Zayas1} to extend the analysis
of \cite{Spence1} for general biHermitian target spaces. 
In \cite{Kapustin3}, the same analysis has been carried out in 
the on--shell formalism. Simultaneously, many mathematical studies
of the problem of reduction of generalized complex, Calabi--Yau and Kaehler
manifolds have appeared \cite{Bursztyn1,Hu,Xu1,Vaisman,Apostolov1,
Tolman1,Tolman2,Lin1,Lin2}, 
calling for a comparison with the target space 
geometries yielded by sigma model gauging.

In this paper, we revisit our earlier work on the AKSZ formulation of 
topological sigma model on generalized complex manifolds, or Hitchin model,  
which we introduced in 2004 in \cite{Zucchini1}. We show that the target 
space geometry geometry encoded in the BV master equations is twisted 
Poisson--quasi--Nijenhuis geometry recently introduced and studied by
Sti\'enon and Xu (in the untwisted case) in \cite{Xu2}. Poisson--quasi--Nijenhuis 
geometry is more general than generalized complex geometry
and comprises it as a particular case. This should clarify
the issue of the underlying geometry of the Hitchin model
raised but not solved in \cite{Zucchini1}. Next, we show how 
gauging (here meant in a non standard way explained in the following)
can be incorporated in the Hitchin model. 
We find that the geometry resulting form the BV master equation 
is closely related to but more general than that described by Lin and Tolman in 
\cite{Tolman1,Tolman2} and is fully $b$ symmetry covariant, suggesting 
a natural framework for the 
study of reduction of twisted Poisson--quasi--Nijenhuis manifolds. 

The plan of the paper is as follows. In sect. \ref{sec:BVparadigm}, 
we review the basic features of the AKSZ formulation of topological sigma
models relevant in the following. In sect. \ref{sec:Weil}, 
we introduce the Weil sigma model, a canonical sigma model associated to 
any real Lie algebra, and study it in the AKSZ framework. 
In sect. \ref{sec:Poisson}, we review the AKSZ formulation of the Poisson 
sigma model and gauge it by coupling it to the Weil model. This introduces
sect. \ref{sec:Hitchin}, where we revisit the AKSZ formulation of the Hitchin 
sigma model showing that the underlying geometry is twisted 
Poisson--quasi--Nijenhuis and gauge it by coupling it again 
to the Weil model. In sect. \ref{sec:geometry}, we study the geometry
of the Hitchin--Weil model and show substantial evidence that this may 
encode a rather general reduction scheme for
Poisson--quasi--Nijenhuis geometry.
Finally, in sect. \ref{sec:remarks}, we discuss the results obtained.

%\vfill\eject

\begin{small}
\section{  \bf The AKSZ paradigm}
\label{sec:BVparadigm}
\end{small}
%\vskip.2cm
The Alexandrov--Kontsevich--Schwartz--Zaboronsky (AKSZ) formalism of ref. \cite{AKSZ} 
is a method of constructing solutions of the Batalin--Vilkovisky (BV) classical 
master equation directly, without starting from a classical action with a 
set of symmetries, as is usually done in the BV framework \cite{BV1,BV2}. 
In ref. \cite{Cattaneo1, Cattaneo2}, using such formalism, Cattaneo and Felder 
managed to obtain the BV action of the Poisson sigma model
\cite{Ikeda2,Strobl}. In spirit, their
approach is essentially the same as the one of the present paper. 
For this reason, we shall review it briefly. We refer the reader to app.
\ref{sec:superfields} for a review of de Rham superfield formalism
used throughout this paper.

Following \cite{Cattaneo2}, we view the standard Poisson sigma model 
as a field theory
whose base space, target space and field configuration space are respectively 
a two dimensional surface $\Sigma$, a Poisson manifold $M$ with Poisson $2$--vector $P$ 
and the space $\mathscri{F}$ of maps $\phi:T[1]\Sigma\mapsto T^*[1]M$. 

The supermanifold $T^*[1]M$ has a canonical odd symplectic structure, or 
$P$--structure, defined by the canonical odd symplectic form $\omega
=du_adt^a$. With $\omega$, there are associated canonical odd Poisson 
brackets $(~,~)_\omega$ in standard fashion.
Indeed, the algebra of functions on $T^*[1]M$ with the odd brackets $(~,~)_\omega$
is isomorphic to the algebra of multivector 
fields on $M$ with the standard Schoutens--Nijenhius brackets. 
The field space $\mathscri{F}$ inherits an odd symplectic structure from that 
of $T^*[1]M$ and, so, it also carries a $P$--structure. 
The associated odd symplectic form $\Omega$ is obtained from $\omega$ 
by integration over $T[1]\Sigma$ with respect to the usual supermeasure $\varrho$ 
(cf. \eqref{measure}).
With $\Omega$, there are associated odd Poisson brackets $(~,~)_\Omega$
over the algebra of functions on the field configuration space $\mathscri{F}$, 
called BV antibrackets in the physical literature.

The base space $T[1]\Sigma$ has a canonical nilpotent odd vector field, or 
$Q$--structure, defined by the usual de Rham differential $d$ on $\Sigma$. $d$ induces 
a $Q$--structure, also denoted by $d$, on the field configuration space 
$\mathscri{F}$ in obvious fashion. $d$ is Hamiltonian, 
as indeed $d=\delta_1=(S_1,)_\Omega$ for a certain function $S_1$ on 
$\mathscri{F}$. $S_1$ satisfies the BV master equation $(S_1,S_1)_\Omega=0$.

The Poisson $2$--vector field $P$ of $M$ can be identified 
with a function on $T^*[1]M$ satisfying $(P,P)_\omega=0$. 
Its Hamiltonian vector field $Q_P=(P,)_\omega$ defines a $Q$ structure 
on $T^*[1]M$. The Poisson $2$--vector $P$ yields a function $S_2$ on 
$\mathscri{F}$, again by integration over $T[1]\Sigma$ with respect to
$\varrho$, satisfying BV master equation $(S_2,S_2)_\Omega=0$. Its
Hamiltonian vector field $\delta_2=(S_2,)_\Omega$ yields in this way
a $Q$--structure on the field configuration space $\mathscri{F}$. 

One verifies that $(S_1,S_2)_\Omega=0$. The sum $S_t=S_1+S_2$ thus satisfies 
the BV master equation $(S_t,S_t)_\Omega=0$. $S_t$ is the BV
action of the Poisson sigma model. Its Hamiltonian vector field 
$\delta_t=(S_t,~)_\Omega$ is the BV variation operator. 

In this paper, we consider sigma models whose base space, target space and 
field configuration space are respectively a two dimensional surface $\Sigma$, 
a supermanifold $X$ carrying various types of algebraic or geometrical structures 
and a space $\mathscri{F}$ of maps $\phi:T[1]\Sigma\mapsto X$. 
A BV odd symplectic form $\Omega$ is defined on $\mathscri{F}$.
$\delta\Omega=0$,
where $\delta$ denotes the de Rham differential in $\mathscri{F}$
\footnote{$\vphantom{\bigg[}$ The $\delta$ should not be confused with the BV
operators $\delta_r$ introduced below.}.
This allows to define BV antibrackets $(~,~)$ on $\mathscri{F}$ in the
usual way. 

The sigma models are characterized by a pair of action functionals 
$S_r$, $r=1,~2$, which satisfy the joined BV master equation
\begin{equation}
(S_r,S_s)=0, \qquad r,~s=1,~2.
\label{BVmaster}
\end{equation}
With the $S_r$ there are associated odd BV variations by
\begin{equation}
\delta_r\phi=(S_r,\phi),
\label{deltardef}
\end{equation}
where $\phi$ is any field of $\mathscri{F}$.
When \eqref{BVmaster} holds, one has
\begin{equation}
\delta_{r}\delta_{s}+\delta_{s}\delta_{r}=0,\qquad r,~s=1,~2. 
\label{}
\end{equation}
Moreover, one has \hphantom{xxxxxxxxxxxxxxxxxxxx}
\begin{equation}
\delta_{r}S_{s}=0,\qquad r,~s=1,~2. 
\label{}
\end{equation}

The general action of the model is of the form
\begin{equation}
S_{t}=t_1S_{1}+t_2S_{2},
\label{}
\end{equation}
where $t\in\mathbb{C}^2\setminus\{0\}$ is a parameter 
\footnote{$\vphantom{\bigg[}$ In certain cases, it may be natural to take $t$ 
to be real. Everything stated below works also under this restriction.}. 
It satisfies the BV master equation
\begin{equation}
(S_t,S_t)=0
\label{}
\end{equation}
The associated BV variation is
\begin{equation}
\delta_t=t_1\delta_{1}+t_2\delta_{2}.
\label{}
\end{equation}
$\delta_t$ is nilpotent,  \hphantom{xxxxxxxxxxxxxxxxxxxx}
\begin{equation}
\delta_t{}^2=0.
\label{}
\end{equation}
Further, one has  \hphantom{xxxxxxxxxxxxxxxxxxxx}
\begin{equation}
\delta_tS_t=0.
\label{}
\end{equation}
We do not consider models with actions functionals differing by an overall 
factor as distinct. So, one actually has a $\mathbb{C}\mathbb{P}^1$
worth of inequivalent models. 

For a given field theory of the type described above, the 
choice of the action functionals $S_r$, $r=1,~2$, is non unique.
One is allowed to carry out a linear redefinition of the form 
\hphantom{xxxxxxxxxxxxxxxxxxxxx}
\begin{equation}
S'{}_r=\sum_{s=1}^2 A_{rs}S_s,
\label{}
\end{equation}
where $(A_{rs})_{r,s=1,2}$ is a non singular $2$ by $2$ complex matrix.
For each sigma model considered in this paper, it is possible to choose 
$S_1$ in such a way that for any field $\phi$, one has  \hphantom{xxxxxxxxxxxxxxxx}
\begin{equation}
\delta_1\phi=d\phi.
\label{}
\end{equation}
%where $d$ is the de Rham differential of the world sheet. 
Upon doing this, $S_2$ is defined up to the addition of a 
complex multiple of $S_1$. 

The similarity of the constructions of this paper with the AKSZ formulation of
the Poisson sigma model of \cite{Cattaneo1,Cattaneo2}
should be manifest now. For this reason we call the above theoretical frame
work the AKSZ paradigm. 

%\vfill\eject

\begin{small}
\section{  \bf The Weil sigma model}
\label{sec:Weil}
\end{small}
In this section, we introduce the Weil sigma model, which plays an 
important role in the following. The Weil model is a canonical sigma model 
associated to any real Lie algebra $\mathfrak{g}$. As it will turn out, 
coupling to the Weil model implements the gauging of the symmetry
associated with the connected Lie group $G$ having $\mathfrak{g}$
as its Lie algebra.

The field content of the model consists of fields 
$\beta\in C^\infty(T[1]\Sigma,\mathfrak{g}^\vee[0])$, 
$\gamma\in C^\infty(T[1]\Sigma,\mathfrak{g}[1])$, 
$\Beta\in C^\infty(T[1]\Sigma,\mathfrak{g}^\vee[-1])$ and
$\Gamma\in C^\infty(T[1]\Sigma,\mathfrak{g}[2])$, 
where $\mathfrak{g}$ is for the time being a real vector space.
The BV odd symplectic form is 
\begin{equation}
\Omega_W=\int_{T[1]\Sigma}\varrho\Big[\delta\beta_i\delta\gamma^i+\delta \Beta_i\delta\Gamma^i\Big].
\end{equation}
This satisfies obviously
\begin{equation}
\delta\Omega_W=0.
\end{equation}
The associated BV rackets are
\begin{equation}
(F,G)_W=\int_{T[1]\Sigma}\varrho\Big[
\frac{\delta_rF}{\delta\beta_i}\frac{\delta_lG}{\delta\gamma^i}
-\frac{\delta_rF}{\delta\gamma^i}\frac{\delta_lG}{\delta\beta_i}
+\frac{\delta_rF}{\delta \Beta_i}\frac{\delta_lG}{\delta\Gamma^i}
-\frac{\delta_rF}{\delta\Gamma^i}\frac{\delta_lG}{\delta \Beta_i}\Big],
\label{Wantbrckt}
\end{equation}
for any two functionals $F$, $G$ on field space.

The model is characterized by two basic action functionals given by 
\begin{subequations}
\begin{align}
S_{W1}&=\int_{T[1]\Sigma}\varrho\Big[\beta_id\gamma^i-\Beta_id\Gamma^i\Big],
\label{}
\\
S_{W2}&=\int_{T[1]\Sigma}\varrho\Big[\beta_i\Gamma^i-\frac{1}{2}f^i{}_{jk}\beta_i\gamma^j\gamma^k
-f^i{}_{jk}\Beta_i\Gamma^j\gamma^k\Big],
\label{}
\end{align}
\end{subequations}
where %the constants $f^i{}_{jk}$ are the components of a tensor 
$f\in \mathfrak{g}\otimes\wedge^2\mathfrak{g}^\vee$.
A simple computation yields the BV brackets
\begin{subequations}
\label{(SW,SW)}
\begin{align}
(S_{W1},S_{W1})_W&=0,
\label{(SW1,SW1)}
\\
(S_{W1},S_{W2})_W&=0,
\label{(SW1,SW2)}
\\
(S_{W2},S_{W2})_W&=2\int_{T[1]\Sigma}\varrho\Big[\frac{1}{6}g^i{}_{jkl}\beta_i\gamma^j\gamma^k\gamma^l
+\frac{1}{2}g^i{}_{jkl}\Beta_i\Gamma^j\gamma^k\gamma^l\Big],
\label{(SW2,SW2)}
\end{align}
\end{subequations}
where $g\in \mathfrak{g}\otimes\wedge^3\mathfrak{g}^\vee$ is given by
\begin{equation}
g^i{}_{jkl}=f^i{}_{mj}f^m{}_{kl}+f^i{}_{mk}f^m{}_{lj}+f^i{}_{ml}f^m{}_{jk}.
\label{gijkl}
\end{equation}
Therefore, the joined BV master equations
\begin{equation}
(S_{Wr},S_{Ws})_W=0, \qquad r,~s=1,~2,
\label{Wmaster}
\end{equation}
are satisfied if and only if  \hphantom{xxxxxxxxxxxx}
\begin{equation}
g^i{}_{jkl}=0,
\label{gijkl=0}
\end{equation}
%with with $g$ given by \eqref{gijkl}, 
that is when $\mathfrak{g}$ is a Lie algebra with structure constants $f^i{}_{jk}$. 
In this way, when \eqref{gijkl=0} is fulfilled, we are in the AKSZ
paradigm described in sect. \ref{sec:BVparadigm}. 

The BV variations associated with the actions $S_{Wr}$ are defined according 
to \eqref{deltardef} as $\delta_{Wr}=(S_{Wr},~)_W$. Explicitly, 
\begin{subequations}
\label{deltaWexp}
\begin{align}
\delta_{W1}\beta_i&=d\beta_i,
\label{deltaW1a}
\\
\delta_{W1}\gamma^i&=d\gamma^i,
\label{deltaW1b}
\\
\delta_{W1}\Beta_i&=d\Beta_i,
\label{deltaW1c}
%\\
\end{align}
\begin{align}
\delta_{W1}\Gamma^i&=d\Gamma^i,
\label{deltaW1d}
\\
\delta_{W2}\beta_i&=-f^j{}_{ik}\beta_j\gamma^k-f^j{}_{ik}\Beta_j\Gamma^k,
\label{deltaW2a}
\\
\delta_{W2}\gamma^i&=\Gamma^i-\frac{1}{2}f^i{}_{jk}\gamma^j\gamma^k,
\label{deltaW2b}
\\
\delta_{W2}\Beta_i&=-\beta_i+f^j{}_{ik}\Beta_j\gamma^k,
\label{deltaW2c}
\\
\delta_{W2}\Gamma^i&=-f^i{}_{jk}\gamma^j\Gamma^k.
\label{deltaW2d}
\end{align}
\end{subequations}

To any Lie algebra $\mathfrak{g}$, there is canonically associated the Weil 
algebra $W(\mathfrak{g})=\wedge^*\mathfrak{g}^\vee[1]\otimes
\vee^*\mathfrak{g}^\vee[2]$. This is the tensor product of 
the antisymmetric and symmetric algebras of $\mathfrak{g}^\vee$
in degree $1$ and $2$, respectively. The natural $\mathfrak{g}$--valued
generators $\omega$, $\Omega$ of $W(\mathfrak{g})$ carry degrees $1$, $2$, 
respectively. The Weil operator $d_W$ acts as
\begin{subequations}
\begin{align}
\label{}
d_W\omega^i&=\Omega^i-\frac{1}{2}f^i{}_{jk}\omega^j\omega^k,
\label{}
\\
d_W\Omega^i&=-f^i{}_{jk}\omega^j\Omega^k,
\label{}
\end{align}
\end{subequations}
and is extended on $W(\mathfrak{g})$ by linearity. $d_W$ is nilpotent
\begin{equation}
d_W{}^2=0.
\label{dW2=0}
\end{equation}
The cohomology of $(W(\mathfrak{g}),d_W)$ is actually trivial
\footnote{$\vphantom{\bigg[}$ As is well known, it is possible to define also a $\mathfrak{g}$ basic
cohomology of $(W(\mathfrak{g}),d_W)$, which turns out to be non trivial.}.
It appears that the fields $\gamma$, $\Gamma$ describe the embedding
of $T[1]\Sigma$ into the Weil algebra. %$C^\infty(T[1]\Sigma,W(\mathfrak{g}))$. 
Further, by \eqref{deltaW2b}, \eqref{deltaW2d}, for any point $z\in T[1]\Sigma$,
the evaluation map $\mathrm{e}_z:C^\infty(T[1]\Sigma,W(\mathfrak{g}))
\mapsto W(\mathfrak{g})$ is a chain map of the chain complexes 
$(C^\infty(T[1]\Sigma,W(\mathfrak{g})),\delta_{W2})$,
$(W(\mathfrak{g}),d_W)$. This justifies the name given to the sigma model
considered above. 

The Weil sigma model describes a supersymmetric gauge ghost system. 
The algebraic structure presented here is closely related to those 
appearing in the so called topological field theories of cohomological type. 
(See sect 10.3 of ref. \cite{Moore1} for a thorough review of these matters
with many illustrative examples).

%\vfill\eject

\begin{small}
\section{  \bf The Poisson--Weil sigma model}
\label{sec:Poisson}
\end{small}

In this section, we illustrate the Poisson--Weil sigma model. This is
interesting on its own and serves also the purpose of introducing the
treatment of the more complicated Hitchin--Weil model expounded later. 
Our presentation is closely related to that of ref. \cite{Zucchini1}, 
in turn inspired by refs. \cite{Cattaneo1,Cattaneo2}.

The field content of the Poisson sigma model consists of 
a degree $0$ embedding $x\in C^\infty(T[1]\Sigma, M)$ and a degree $1$
section $y\in C^\infty(T[1]\Sigma, x^*T^*[1]M)$. 
The BV odd symplectic form is \hphantom{xxxxxxxxxxxx}
\begin{equation}
\Omega_M=\int_{T[1]\Sigma}\varrho\,\delta x^a\delta y_a.
\end{equation}
This satisfies obviously  \hphantom{xxxxxxxxxxxx}
\begin{equation}
\delta\Omega_M=0.
\end{equation}
The associated BV antibrackets are given by
\begin{equation}
(F,G)_M=\int_{T[1]\Sigma}\varrho\,\bigg[
\frac{\delta_r F}{\delta x^a}\frac{\delta_l G}{\delta y_a}-
\frac{\delta_r F}{\delta y_a}\frac{\delta_l G}{\delta x^a}\bigg],
\label{Mantbrckt}
\end{equation}
for any two functionals $F$, $G$ on field space. See app. \ref{sec:techno} 
for technical details.

The model is characterized by two action functionals
\begin{subequations}
\begin{align}
S_{P1}&=\int_{T[1]\Sigma}\varrho \,y_adx^a,
\label{}
\\
S_{P2}&=\int_{T[1]\Sigma}\varrho \,\frac{1}{2}P^{ab}(x)y_ay_b,
\label{}
\end{align}
\end{subequations}
where $P\in C^\infty(M,\wedge^2TM)$ is a $2$--vector defining 
an almost Poisson structure on $M$.

A simple computation yields the BV brackets
\begin{subequations}
\label{(SP,SP)}
\begin{align}
(S_{P1},S_{P1})_M&=0,
\label{(SP1,SP1)}
\\
(S_{P1},S_{P2})_M&=0,
\label{(SP1,SP2)}
\\
(S_{P2},S_{P2})_M&=2\int_{T[1]\Sigma}\varrho\,\Big[-\frac{1}{6}A^{abc}(x)y_ay_by_c\Big], 
\label{(SP2,SP2)}
\end{align}
\end{subequations}
where the $3$--vector $A\in C^\infty(M, \wedge^3TM)$ is given by
\begin{equation}
A^{abc}=P^{ad}\partial_dP^{bc}+P^{bd}\partial_dP^{ca}+P^{cd}\partial_dP^{ab}.
\label{Aabc}
\end{equation} 
Therefore, the joined BV master equations
\begin{equation}
(S_{Pr},S_{Ps})_M=0, \qquad r,~s=1,~2,
\label{Pmaster}
\end{equation}
are satisfied if and only if  \hphantom{xxxxxxxxxxxxxxx}
\begin{equation}
A^{abc}=0.
\label{Aabc=0}
\end{equation}
%with $A$ given by \eqref{Aabc}. 
In this way, when \eqref{Aabc=0} holds, we are 
in the AKSZ paradigm described in sect. \ref{sec:BVparadigm}. 
As is well--known,
condition \eqref{Aabc=0} ensures the almost Poisson structure $P$ is 
actually a Poisson structure, so that $M$ is a Poisson manifold. 

The BV variations associated with the actions $S_{Pr}$ are defined according 
to \eqref{deltardef} as $\delta_{Pr}=(S_{Pr},~)_M$. Explicitly, one has
\begin{subequations}
\label{deltaPexp}
\begin{align}
\delta_{P1} x^a&=dx^a,
\label{deltaP1a}
\\
\delta_{P1} y_a&=dy_a,
\label{deltaP1b}
\\
\delta_{P2} x^a&=P^{ab}(x)y_b,
\label{deltaP2a}
\\
\delta_{P2} y_a&=\frac{1}{2}\partial_aP^{bc}(x)y_by_c.
\label{deltaP2b}
\end{align}
\end{subequations}

One can couple the Poisson and the Weil sigma models to obtain the 
Poisson--Weil sigma model. The field space of Poisson--Weil model  
is simply the Cartesian product of those of the Poisson and Weil models.  
The BV odd symplectic form $\Omega_{MW}$ of the Poisson--Weil model 
is correspondingly the sum of those of the Poisson and Weil models, 
$\Omega_{MW}=\Omega_M+\Omega_W$.
%\begin{equation}
%\label{OmegaPW}
%\end{equation}
Consequently, the BV antibrackets $(~,~)_{MW}$ are the sum of the BV 
antibrackets $(~,~)_M$ and $(~,~)_W$ given by \eqref{Mantbrckt}, 
\eqref{Wantbrckt}.

The Poisson--Weil model is characterized by two action functionals:
\begin{subequations}
\begin{align}
S_{PW1}&=S_{P1}+S_{W1},
\label{}
\\
S_{PW2}&=S_{P2}+S_{W2}+\int_{T[1]\Sigma}\varrho\Big[-u_i{}^a(x)\gamma^iy_a+\mu_i(x)\Gamma^i\Big],
\label{}
\end{align}
\end{subequations}
where $u\in C^\infty(M,\mathfrak{g}^\vee \otimes TM)$ and $\mu\in C^\infty(M,\mathfrak{g}^\vee)$ 
are a $\mathfrak{g}^\vee$--valued vector field and a $\mathfrak{g}^\vee$--valued scalar on
$M$, respectively. 

A straightforward computation yields the BV brackets
\begin{subequations}
\begin{align}
(S_{PW1},S_{PW1})_{MW}&=0,
\label{}
\\
(S_{PW1},S_{PW2})_{MW}&=0,
\label{}
\\
(S_{PW2},S_{PW2})_{MW}&=(S_{P2},S_{P2})_M+(S_{W2},S_{W2})_W
\\
\label{}
+2\int_{T[1]\Sigma}\varrho\,\Big[
\frac{1}{2}X_i&{}^{ab}(x)\gamma^iy_ay_b
-\frac{1}{2}L_{ij}{}^a(x)\gamma^i\gamma^jy_a
+N_{ij}(x)\gamma^i\Gamma^j-S_i{}^a(x)\Gamma^iy_a
\Big], 
\nonumber
\end{align}
\end{subequations}
where the BV antibrackets $(S_{P2},S_{P2})_M$, $(S_{W2},S_{W2})_W$ are given
by \eqref{(SP2,SP2)}, \eqref{(SW2,SW2)}, respectively, and 
$X\in C^\infty(M,\mathfrak{g}^\vee\otimes\wedge^2TM)$,
$L\in C^\infty(M,\wedge^2\mathfrak{g}^\vee\otimes TM)$,
$N\in C^\infty(M,\mathfrak{g}^\vee\otimes\mathfrak{g}^\vee)$, 
$S\in C^\infty(M,\mathfrak{g}^\vee\otimes TM)$ are given by  
\begin{subequations}
\label{PWdefs}
\begin{align}
X_i{}^{ab}&=u_i{}^c\partial_cP^{ab}-\partial_cu_i{}^aP^{cb}-\partial_cu_i{}^bP^{ac},
\label{PWdefsa}
\\
L_{ij}{}^a&=u_i{}^b\partial_bu_j{}^a-u_j{}^b\partial_bu_i{}^a-f^k{}_{ij}u_k{}^a,
\label{PWdefsb}
\\
N_{ij}{}&=u_i{}^b\partial_b\mu_j-f^k{}_{ij}\mu_k,
\label{PWdefcs}
\\
S_i{}^a&=u_i{}^a+P^{ab}\partial_b\mu_i.
\label{PWdefsd}
\end{align}
\end{subequations}
The joined BV master equations
\begin{equation}
(S_{PWr},S_{PWs})_{MW}=0, \qquad r,~s=1,~2,
\label{PWmaster}
\end{equation}
are satisfied if and only if \eqref{Aabc=0}, the conditions 
\begin{subequations}
\label{PWmastercnds}
\begin{align}
N_{ij}&=0,
\label{PWmastercndsc}
\\
S_i{}^a&=0,
\label{PWmastercndsd}
\end{align}
\end{subequations}
and \eqref{gijkl=0} are simultaneously fulfilled.
Indeed, it is easy to see that, when $u_i$ is given by \eqref{PWmastercndsd}, 
one has
\begin{subequations}
\label{PWrels}
\begin{align}
X_i{}^{ab}&=A^{abc}\partial_c\mu_i,
\label{PWrelsa}
\\
L_{ij}{}^a&=A^{abc}\partial_b\mu_i\partial_c\mu_j-P^{ab}\partial_bN_{ij},
\label{PWrelsb}
\end{align}
\end{subequations}
In this way, when \eqref{Aabc=0}, \eqref{PWmastercnds} and \eqref{gijkl=0}
hold, we are again in the AKSZ paradigm described in sect. \ref{sec:BVparadigm}.
The geometry of $M$ emerging here will be analyzed in greater detail  
in sect. \ref{sec:geometry}. We anticipate that 
that $M$ is a Poisson manifold carrying an infinitesimal 
action of the Lie algebra $\mathfrak{g}$ leaving the Poisson structure $P$ 
invariant, the action being Hamiltonian with equivariant moment map $\mu$.
This geometrical set up allows for the symmetry reduction of $M$, which
is therefore encoded in the Poisson--Weil model. 

The BV variations associated with the actions $S_{PWr}$ are defined as usual
according to \eqref{deltardef} as $\delta_{PWr}=(S_{PWr},~)_{MW}$. Explicitly,
one has
\begin{subequations}
\begin{align}
\delta_{PW1} x^a&=\delta_{P1} x^a,
\label{}
\\
\delta_{PW1} y_a&=\delta_{P1} y_a,
\label{}
\\
\delta_{PW1}\beta_i&=\delta_{W1}\beta_i,
\\
\delta_{PW1}\gamma^i&=\delta_{W1}\gamma^i,
\label{}
\\
\delta_{PW1}\Beta_i&=\delta_{W1}\Beta_i,
\label{}
\\
\delta_{PW1}\Gamma^i&=\delta_{W1}\Gamma^i,
\label{}
\\
\delta_{PW2} x^a&=\delta_{P2} x^a+u_i{}^a(x)\gamma^i,
\label{}
\\
\delta_{PW2} y_a&=\delta_{P2} y_a-\partial_au_i{}^b(x)\gamma^iy_b+\partial_a\mu_i(x)\Gamma^i,
\label{}
\\
\delta_{PW2}\beta_i&=\delta_{W2}\beta_i-u_i{}^a(x)y_a,
\\
\delta_{PW2}\gamma^i&=\delta_{W2}\gamma^i,
\label{}
\\
\delta_{PW2}\Beta_i&=\delta_{W2}\Beta_i-\mu_i(x),
\label{}
\\
\delta_{PW2}\Gamma^i&=\delta_{W2}\Gamma^i,
\label{}
\end{align}
\end{subequations}
where the variations $\delta_{Pr}$, $\delta_{Wr}$ are given in 
\eqref{deltaPexp}, \eqref{deltaWexp}, respectively.

%\vfill\eject

\begin{small}
\section{  \bf The Hitchin--Weil sigma model}
\label{sec:Hitchin}
\end{small}

In this section, we illustrate the Hitchin--Weil sigma model, which is the main
topic of this paper. We follow closely the AKSZ treatment of ref. \cite{Zucchini1}.
This will lead us on one hand to realize that the underlying geometry of the model
is Poisson--quasi--Nijenhuis rather than generalized complex, on the other it
will give us useful indications about symmetry reduction in this context, to
be discussed in detail in sect. \ref{sec:geometry}. 

The target space of the Hitchin sigma model is a twisted manifold, i. e. a manifold 
$M$ equipped with a closed $3$--form $H\in C^\infty(M,\wedge^3T^*M)$,
\footnote{$\vphantom{\bigg[}$ The sign convention of the $H$ field used here is opposite to that
employed in ref. \cite{Zucchini1}.}
\begin{equation}
\partial_aH_{bcd}-\partial_bH_{acd}+\partial_cH_{abd}-\partial_dH_{abc}=0.
\label{dH=0}
\end{equation}
The field content of the Hitchin sigma model consists of a degree $0$
embedding $x\in C^\infty(T[1]\Sigma, M)$ and a degree $1$
section $y\in C^\infty(T[1]\Sigma, x^*T^*[1]M)$ as for the Poisson sigma model. 
The BV odd symplectic form is 
\begin{equation}
\Omega_{M,H}=\int_{T[1]\Sigma}\varrho\,\Big[\delta x^a\delta y_a
-\frac{1}{2}H_{abc}(x)\delta x^a dx^b \delta x^c\Big].
\label{OmegaMH}
\end{equation}
It is easy to check that $\Omega_{M,H}$ satisfies 
\begin{equation}
\delta\Omega_{M,H}=0
\end{equation}
on account of \eqref{dH=0}. The associated BV antibrackets are given by
\begin{equation}
(F,G)_{M,H}=\int_{T[1]\Sigma}\varrho\,\bigg[
\frac{\delta_r F}{\delta x^a}\frac{\delta_l G}{\delta y_a}-
\frac{\delta_r F}{\delta y_a}\frac{\delta_l G}{\delta x^a}
+H_{abc}(x)\frac{\delta_r F}{\delta y_a} 
dx^b \frac{\delta_l G}{\delta y_c}\bigg],
\label{MHantbrckt}
\end{equation}
for any two functionals $F$, $G$ on field space. See again
app. \ref{sec:techno} for technical details.

The model is characterized by two action functionals
\begin{subequations}
\begin{align}
S_{H1}&=\int_{T[1]\Sigma}\varrho \,y_adx^a+2\int_\Gamma x^{(0)*}H,
\label{}
\\
S_{H2}&=\int_{T[1]\Sigma}\varrho \,
\Big[\frac{1}{2}P^{ab}(x)y_ay_b+J^a{}_b(x)y_adx^b\Big]+\int_\Gamma x^{(0)*}\Phi.
%+\frac{1}{2}Q_{ab}(x)dx^adx^b\Big],
\label{}
\end{align}
\end{subequations}
Here, $\Gamma$ is a $3$--fold such that $\partial \Gamma=\Sigma$ and 
$x^{(0)}:\Gamma\rightarrow M$ is an embedding such that $x^{(0)}|_\Sigma$
equals the lowest degree $0$ component of the embedding superfield $x$ 
(see app. \ref{sec:superfields}) and 
whose choice is immaterial. $P\in C^\infty(M,\wedge^2TM)$, 
$J\in C^\infty(M,\mathrm{End}\,TM)$, $\Phi\in C^\infty(M,\wedge^3T^*M)$, 
are respectively a $2$--vector, an endomorphism and a closed $3$--form 
\hphantom{xxxxxxxxxxxxxxx}
\begin{equation}
\partial_a\Phi_{bcd}-\partial_b\Phi_{acd}+\partial_c\Phi_{abd}-\partial_d\Phi_{abc}=0.
\label{dPhi=0}
\end{equation}
Further, the compatibility condition
% defining a generalized almost complex structure
\begin{equation}
J^a{}_cP^{cb}+J^b{}_cP^{ca}=0
\label{JPcomp}
\end{equation}
holds. The tensors $P$, $J$ and $\Phi$ together define 
an almost Poisson--quasi--Nijenhuis structure \cite{Xu2}. 
The version of the Hitchin model presented here is 
more general than that originally expounded in 
\cite{Zucchini1}, where the $3$--form $\Phi$ was 
assumed to be exact (cf. eq. \eqref{Phi=dQ} below). 

A straightforward computation yields the BV brackets
\begin{subequations}
\label{(SH,SH)}
\begin{align}
(S_{H1},S_{H1})_{M,H}&=0,
\label{(SH1,SH1)}
\\
(S_{H1},S_{H2})_{M,H}&=0,
\label{(SH1,SH2)}
\\
(S_{H2},S_{H2})_{M,H}&=2\int_{T[1]\Sigma}\varrho\,\Big[
-\frac{1}{6}A_H{}^{abc}(x)y_ay_by_c
\label{(SH2,SH2)}
\\
&~~~~~~~~~~~~~~~~~~+\frac{1}{2}B_H{}^{ab}{}_c(x)y_ay_bdx^c
-\frac{1}{2}C_H{}^a{}_{bc}(x)y_adx^bdx^c\Big], 
\nonumber
\end{align}
\end{subequations}
where the tensor $A_H\in C^\infty(M, \wedge^3TM)$,
$B_H\in C^\infty(M,\wedge^2TM\otimes T^*M)$,
$C_H\in C^\infty(M,TM\otimes \wedge^2T^*M)$
are given by
\begin{subequations}
\label{AHBHCH}
\begin{align}
A_H{}^{abc}&=P^{ad}\partial_dP^{bc}+P^{bd}\partial_dP^{ca}+P^{cd}\partial_dP^{ab},
\label{AHabc}
\\
B_H{}^{ab}{}_c&=J^d{}_c\partial_dP^{ab}
+P^{ad}(\partial_cJ^b{}_d-\partial_d J^b{}_c)
-P^{bd}(\partial_cJ^a{}_d -\partial_dJ^a{}_c)
\label{BHabc}
\\
&~~~~~~-\partial_c(J^a{}_dP^{db})-P^{ad}P^{be}H_{cde},
\nonumber
\\
C_H{}^a{}_{bc}&=J^d{}_b\partial_dJ^a{}_c-J^d{}_c\partial_dJ^a{}_b
-J^a{}_d\partial_bJ^d{}_c+J^a{}_d\partial_cJ^d{}_b
\label{CHabc}
\\
&~~~~~~+P^{ad}\Phi_{dbc}   %(\partial_dQ_{bc}+\partial_bQ_{cd}+\partial_cQ_{db})
+J^d{}_bP^{ae}H_{cde}-J^d{}_cP^{ae}H_{bde}.
\nonumber
\end{align}
\end{subequations}
Therefore, the joined BV master equations
\begin{equation}
(S_{Hr},S_{Hs})_{M,H}=0, \qquad r,~s=1,~2,
\label{Hmaster}
\end{equation}
are satisfied if and only if 
\begin{subequations}
\label{PqNconds}
\begin{align}
A_H{}^{abc}&=0,
\label{AHabc=0}
\\
B_H{}^{ab}{}_c&=0,
\label{BHabc=0}
\\
C_H{}^a{}_{bc}&=0.
\label{CHabc=0}
\end{align}
\end{subequations}
%with $A_H$, $B_H$, $C_H$  given by \eqref{AHBHCH}. 
In this way, when \eqref{PqNconds} holds, we are in the AKSZ
paradigm described in sect. \ref{sec:BVparadigm}. 
Conditions \eqref{PqNconds} are satisfied when 
%states the conditions which render 
the almost Poisson--quasi--Nijenhuis structure $(P,J,\Phi)$ is 
an $H$--twisted Poisson--quasi--Nijenhuis structure.
A more restrictive notion of Poisson--quasi--Nijenhuis manifold was introduced by 
Sti\'enon and Xu in \cite{Xu2} in the untwisted case $H=0$ (see
sect. \ref{sec:geometry} below). As appears, the target space geometry of the
Hitchin model encoded in the BV master
equations is twisted Poisson--quasi--Nijenhuis. This broadens the scope of our
original work on this model \cite{Zucchini1}. (See also \cite{Ikeda1,Ikeda3}
for an alternative approach).

Twisted generalized complex geometry is a special case of twisted
Poisson--quasi-Nijenhuis geometry. For a generalized almost complex manifold,
the $3$--form $\Phi$ is exact, so that one has  
\begin{equation}
\Phi_{abc}=\partial_aQ_{bc}+\partial_bQ_{ca}+\partial_cQ_{ab}, 
\label{Phi=dQ}
\end{equation}
for some $Q\in C^\infty(M,\wedge^2T^*M)$. 
The compatibility conditions are \eqref{JPcomp} and 
\begin{subequations}
\label{JPQcomp}
\begin{align}
&J^a{}_cJ^c{}_b+P^{ac}Q_{cb}+\delta^a{}_b=0,
\label{JPQcompa}
\\
&Q_{ac}J^c{}_b+Q_{bc}J^c{}_a=0.
\label{JPQcompb}
\end{align}
\end{subequations}
The differential conditions \eqref{PqNconds} are necessary but not sufficient  
for the target space generalized almost complex structure to be
Courant integrable. To have Courant integrability, one needs, besides
\eqref{PqNconds}, a further condition
\begin{equation}
D_{Habc}=0
\label{DH=0}
\end{equation}
where $D_H\in C^\infty(M, \wedge^3T^*M)$ is a $3$--form defined by
\begin{align}
D_H{}_{abc}&=J^d{}_a\Phi_{dbc}      %(\partial_dQ_{bc}+\partial_bQ_{cd}+\partial_cQ_{db})
+J^d{}_b \Phi_{dca}        %(\partial_dQ_{ca}+\partial_cQ_{ad}+\partial_aQ_{dc})
+J^d{}_c\Phi_{dab}        %(\partial_dQ_{ab}+\partial_aQ_{bd}+\partial_bQ_{da})
-\partial_a(Q_{bd}J^d{}_c)-\partial_b(Q_{cd}J^d{}_a)
\label{DH}
\\
&~~~~~~-\partial_c(Q_{ad}J^d{}_b)
+H_{abc}-J^d{}_aJ^e{}_bH_{cde}-J^d{}_bJ^e{}_cH_{ade}-J^d{}_cJ^e{}_aH_{bde}.
\nonumber
\end{align}
The Courant integrability conditions \eqref{PqNconds}, \eqref{DH=0} were first derived in
\cite{Lindstrom2} and in equivalent form in \cite{Zucchini1} before 
Poisson--quasi--Nijenhuis geometry was formulated in \cite{Xu2}. 

The BV variations associated with the actions $S_{Hr}$ are defined according 
to \eqref{deltardef} as $\delta_{Hr}=(S_{Hr},~)_{M,H}$. Explicitly, one has
\begin{subequations}
\label{deltaHexp}
\begin{align}
\delta_{H1} x^a&=dx^a,
\label{deltaH1a}
\\
\delta_{H1} y_a&=dy_a,
\label{deltaH1b}
\\
\delta_{H2} x^a&=P^{ab}(x)y_b+J^a{}_b(x)dx^b,
\label{deltaH2a}
\\
\delta_{H2} y_a&=\frac{1}{2}\partial_aP^{bc}(x)y_by_c
+(\partial_aJ^b{}_c-\partial_cJ^b{}_a-P^{bd}H_{dac})(x)y_bdx^c
\label{deltaH2b}
\\
&+J^b{}_a(x)dy_b+\frac{1}{2}(\Phi_{abc}      %\partial_aQ_{bc}+\partial_bQ_{ca}+\partial_cQ_{ab}
-J^d{}_cH_{abd}+J^d{}_bH_{acd})(x)dx^bdx^c
\nonumber
\end{align}
\end{subequations}

One can couple the Hitchin and the Weil sigma models and obtain the Hitchin--Weil 
sigma model, as one did for the Poisson sigma model. The field space of 
Hitchin--Weil model is simply the Cartesian product of those of the Hitchin 
and Weil models. The BV odd symplectic form $\Omega_{MW,H}$  of the Hitchin--Weil 
model is correspondingly the sum of those of the Hitchin and Weil models,
$\Omega_{MW,H}=\Omega_{M,H}+\Omega_W$.
%\begin{equation}
%\label{OmegaPW}
%\end{equation}
The BV antibrackets $(~,~)_{MW,H}$ are simply the sum of the BV 
antibrackets $(~,~)_{M,H}$ and $(~,~)_W$ given by \eqref{MHantbrckt}, 
\eqref{Wantbrckt}.

The Hitchin--Weil model is characterized by two action functionals,
\begin{subequations}
\begin{align}
S_{HW1}&=S_{H1}+S_{W1},
\label{}
\\
S_{HW2}&=S_{H2}+S_{W2}+\int_{T[1]\Sigma}\varrho\Big[
i\beta_id\gamma^i-i\Beta_id\Gamma^i-u_i{}^a(x)\gamma^iy_a
\label{}
\\
&\hskip 5cm -(\tau_{ia}-i\partial_a\mu_i)(x)\gamma^idx^a+\mu_i(x)\Gamma^i\Big],
\nonumber
\end{align}
\end{subequations}
where $u\in C^\infty(M,\mathfrak{g}^\vee\otimes TM)$,
$\tau\in C^\infty(M,\mathfrak{g}^\vee \otimes T^*M)$ 
and $\mu\in C^\infty(M,\mathfrak{g}^\vee)$ 
are a $\mathfrak{g}^\vee$--valued vector field, a $\mathfrak{g}^\vee$--valued 
$1$--form and a $\mathfrak{g}^\vee$--valued scalar on $M$, respectively.
We note that the action $S_{HW2}$ is intrinsically complex because of the
factors $i$ appearing in the third term. 
%Below we assume that the $3$--form $\Phi$ is exact, so that \eqref{Phi=dQ} holds.

The computation of the BV brackets of the $S_{HWr}$ is lengthy but completely 
straightforward. The result is 
\begin{subequations}
\begin{align}
&(S_{HW1},S_{HW1})_{MW,H}=0,\vphantom{\frac{1}{2}}\hskip6cm
\label{}
\\
&(S_{HW1},S_{HW2})_{MW,H}=0,\vphantom{\frac{1}{2}}
\label{}
\\
&(S_{HW2},S_{HW2})_{MW,H}=(S_{H2},S_{H2})_M+(S_{W2},S_{W2})_W\vphantom{\frac{1}{2}}
\\
\label{}
&\hskip .5cm +2\int_{T[1]\Sigma}\varrho\,\Big[
\frac{1}{2}X_i{}^{ab}(x)\gamma^iy_ay_b
+Y_i{}^a{}_b(x)\gamma^iy_adx^b
+\frac{1}{2}Z_{iab}(x)\gamma^idx^adx^b
\nonumber
\\
&\hskip 2.4cm-\frac{1}{2}L_{ij}{}^a(x)\gamma^i\gamma^jy_a-\frac{1}{2}M_{ija}(x)\gamma^i\gamma^jdx^a
+N_{ij}(x)\gamma^i\Gamma^j
\nonumber
\\
&\hskip 2.4cm-R_{ij}(x)\gamma^id\gamma^j-S_i{}^a(x)\Gamma^iy_a-T_{ia}(x)\Gamma^idx^a
+V_i{}^a(x)d\gamma^iy_a\vphantom{\frac{1}{2}}\Big], 
\nonumber
\end{align}
\end{subequations}
where the BV antibrackets $(S_{H2},S_{H2})_M$, $(S_{W2},S_{W2})_W$ are given
by \eqref{(SH2,SH2)}, \eqref{(SW2,SW2)}, respectively, and 
$X\in C^\infty(M,\mathfrak{g}^\vee\otimes\wedge^2TM)$,
$Y\in C^\infty(M,\mathfrak{g}^\vee\otimes\mathrm{End}\,TM)$,
$Z\in C^\infty(M,\mathfrak{g}^\vee\otimes\wedge^2T^*M)$,
$L\in C^\infty(M,\wedge^2\mathfrak{g}^\vee\otimes TM)$,
$M\in C^\infty(M,\wedge^2\mathfrak{g}^\vee\otimes T^*M)$,
$N,R\in C^\infty(M,\mathfrak{g}^\vee\otimes\mathfrak{g}^\vee)$,
$S,V\in C^\infty(M,\mathfrak{g}^\vee\otimes TM)$,
$T\in C^\infty(M,\mathfrak{g}^\vee\otimes T^*M)$
are given by  
\begin{subequations}
\label{HWdefs}
\begin{align}
X_i{}^{ab}&=u_i{}^c\partial_cP^{ab}-\partial_cu_i{}^aP^{cb}-\partial_cu_i{}^bP^{ac},
\label{HWdefsa}
\\
Y_i{}^a{}_b&=u_i{}^c\partial_cJ^a{}_b-\partial_cu_i{}^aJ^c{}_b+\partial_bu_i{}^cJ^a{}_c
-P^{ac}\Upsilon_{icb},
\label{HWdefsb}
\\
Z_{iab}&=u_i{}^c\Phi_{cab}-\partial_a\Xi_{ib}+\partial_b\Xi_{ia}
+J^c{}_a\Upsilon_{icb}-J^c{}_b\Upsilon_{ica},
\label{HWdefsc}
\\
L_{ij}{}^a&=u_i{}^b\partial_bu_j{}^a-u_j{}^b\partial_bu_i{}^a-f^k{}_{ij}u_k{}^a,
\label{HWdefsd}
\\
M_{ija}&=\frac{1}{2}\Big[u_i{}^b\partial_b\tau_{ja}+\partial_au_i{}^b\tau_{jb}
-u_j{}^b\partial_b\tau_{ia}-\partial_au_j{}^b\tau_{ib}-2f^k{}_{ij}\tau_{ka}
\label{HWdefse}
\\
&\hskip 1cm-u_j{}^b\Upsilon_{iba}+u_i{}^b\Upsilon_{jba}
-i\partial_a(u_i{}^b\partial_b\mu_j-u_j{}^b\partial_b\mu_i-2f^k{}_{ij}\mu_k)\Big],
\nonumber
\\
N_{ij}{}&=u_i{}^a\partial_a\mu_j-f^k{}_{ij}\mu_k,
\label{HWdefsf}
\\
R_{ij}{}&=\frac{1}{2}\Big[u_i{}^a\tau_{ja}+u_j{}^a\tau_{ia}
-i(u_i{}^a\partial_a\mu_j+u_j{}^a\partial_a\mu_i)\Big],
\label{HWdefsg}
\\
S_i{}^a&=u_i{}^a+P^{ab}\partial_b\mu_i,
\label{HWdefsh}
\\
T_{ia}&=\tau_{ia}-J^b{}_a\partial_b\mu_i,
\label{HWdefsi}
\\
V_i{}^a&=J^a{}_bu_i{}^b+P^{ab}(\tau_{ib}-i\partial_b\mu_i)-iu_i{}^a,
\label{HWdefsj}
\end{align}
where $\Xi\in C^\infty(M,\mathfrak{g}^\vee\otimes T^*M)$, 
$\Upsilon\in C^\infty(M,\mathfrak{g}^\vee\otimes \wedge^2T^*M)$
are given by 
\begin{align}
\Xi_{ia}&=i(\delta^b{}_a-iJ^b{}_a)(\tau_{ib}-i\partial_b\mu_i),
\label{HWdefsk}
\\
\Upsilon_{iab}&=\partial_a\tau_{ib}-\partial_b\tau_{ia}-u_i{}^cH_{cab}.
\label{HWdefsl}
\end{align}
\end{subequations}
The joined BV master equations
\begin{equation}
(S_{HWr},S_{HWs})_{MW,H}=0, \qquad r,~s=1,~2,
\label{HWmaster}
\end{equation}
are satisfied if and only if \eqref{PqNconds}, the conditions
\begin{subequations}
\label{HWmastercnds}
\begin{align}
N_{ij}&=0,
\label{HWmastercndsf}
\\
S_i{}^a&=0,
\label{HWmastercndsh}
\\
T_{ia}&=0,
\label{HWmastercndsi}
\end{align}
\end{subequations}
and \eqref{gijkl=0} are simultaneously fulfilled. 
Indeed, it is not difficult to check that, 
when $u_i$ and $\tau_i$ are given by \eqref{HWmastercndsh}
and \eqref{HWmastercndsi}, respectively, one has
\begin{subequations}
\label{HWrels}
\begin{align}
X_i{}^{ab}&=A_H{}^{cab}\partial_c\mu_i,
\label{HWrelsa}
\\
Y_i{}^a{}_b&=-B_H{}^{ca}{}_b\partial_c\mu_i,
\label{HWrelsb}
\\
Z_{iab}&=C_H{}^c{}_{ab}\partial_c\mu_i
\label{HWrelsc}
\\
L_{ij}{}^a&=A_H{}^{abc}\partial_b\mu_i\partial_c\mu_j-P^{ab}\partial_bN_{ij},
\label{HWrelsd}
\\
M_{ija}&=-B_H{}^{bc}{}_a\partial_b\mu_i\partial_c\mu_j-i(\delta^b{}_a+iJ^b{}_a)\partial_bN_{ij},
\\
R_{ij}&=0,
\label{HWrelse}
\\
V_i{}^a&=0.
\label{HWrelsf}
\end{align}
\end{subequations}
In this way,  when \eqref{PqNconds}, \eqref{HWmastercnds} and \eqref{gijkl=0} 
hold, we are again in the AKSZ paradigm described in sect. \ref{sec:BVparadigm}. 
The geometrical interpretation of conditions %\eqref{PqNconds}, 
\eqref{HWmastercnds} 
will be analyzed later in sect. \ref{sec:geometry}. We anticipate that 
the geometry they describe is closely related to but more general than
that of reduction of generalized complex and Kaehler manifolds under a 
group action recently developed by Lin and Tolman in \cite{Tolman1,Tolman1}
and  may suggest a viable framework for reduction of 
Poisson--quasi--Nijenhuis manifolds. 

In the formulation of refs. \cite{Tolman1,Tolman1}, generalized complex 
geometry being concerned, \eqref{Phi=dQ}--\eqref{DH=0} hold true. 
In addition 
to \eqref{PqNconds}, \eqref{HWmastercnds} and \eqref{gijkl=0}, it is further 
assumed that \hphantom{xxxxxxxxxxxxxxxxxxxxxxxxx}
\begin{equation}
\Upsilon_{ia}=0,
\label{LTcnds}
\end{equation}
where $\Upsilon$ is given by \eqref{HWdefsl}. All the tensors appearing
in \eqref{HWrels} continue of course to vanish, but one also has 
a further relation, which pairs with \eqref{HWrelsf},
\begin{equation}
W_{ia}=0,
\label{Wi=0}
\end{equation}
where $W\in C^\infty(M, \mathfrak{g}^\vee\otimes T^*M)$ is given by  
\begin{equation}
W_{ia}=Q_{ab}u_i{}^b-J^b{}_a(\tau_{ib}-i\partial_b\mu_i)-i(\tau_{ia}-i\partial_a\mu_i). 
\end{equation}
These conditions plus other regularity conditions are sufficient to ensure 
the existence of a reduction of the relevant generalized complex manifold.

The BV variations associated with the actions $S_{HWr}$ are defined as usual
according to \eqref{deltardef} as $\delta_{HWr}=(S_{HWr},~)_{MW,H}$. Explicitly,
\begin{subequations}
\begin{align}
\delta_{HW1} x^a&=\delta_{H1} x^a,
\label{}
\\
\delta_{HW1} y_a&=\delta_{H1} y_a,
\label{}
\\
\delta_{HW1}\beta_i&=\delta_{W1}\beta_i,
\\
\delta_{HW1}\gamma^i&=\delta_{W1}\gamma^i,
\label{}
\\
\delta_{HW1}\Beta_i&=\delta_{W1}\Beta_i,
\label{}
\\
\delta_{HW1}\Gamma^i&=\delta_{W1}\Gamma^i,
\label{}
\\
\delta_{HW2} x^a&=\delta_{H2}x^a+u_i{}^a(x)\gamma^i,
\label{}
\\
\delta_{HW2} y_a&=\delta_{H2}y_a-\partial_au_i{}^b(x)\gamma^iy_b
-(\tau_{ia}-i\partial_a\mu_i)(x)d\gamma^i
\label{}
\\
&\hskip .5 cm -(\partial_a\tau_{ib}-\partial_b\tau_{ia}-u_i{}^cH_{cab})(x)\gamma^idx^b
+\partial_a\mu_i(x)\Gamma^i,
\nonumber
\\
\delta_{HW2}\beta_i&=\delta_{W2}\beta_i+id\beta_i
-u_i{}^a(x)y_a-(\tau_{ia}-i\partial_a\mu_i)(x)dx^a,
%\\
\end{align}
\begin{align}
\delta_{HW2}\gamma^i&=\delta_{W2}\gamma^i+id\gamma^i,\hskip5.5cm
\label{}
\\
\delta_{HW2}\Beta_i&=\delta_{W2}\Beta_i+id\Beta_i-\mu_i(x),
\label{}
\\
\delta_{HW2}\Gamma^i&=\delta_{W2}\Gamma^i+id\Gamma^i,
\label{}
\end{align}
\end{subequations}
where the variations $\delta_{Hr}$, $\delta_{Wr}$ are given in 
\eqref{deltaHexp}, \eqref{deltaWexp}, respectively.

$b$ transformation is the basic symmetry of generalized complex geometry.
Though originally discovered in this context, $b$ transformation can be 
straightforwardly generalized to twisted Poisson--quasi--Nijenhuis geometry.
For a thorough analysis of the significance of $b$ transformation, the reader 
is referred to \cite{Gualtieri}. 

$b$ transformation is parameterized by a $2$--form $b\in
C^\infty(M,\wedge^2T^*M)$. It acts in the $3$--form $H$ by shifting it
by $d_Mb$:
\begin{equation} 
H'{}_{abc}=H_{abc}+\partial_ab_{bc}+\partial_bb_{ca}+\partial_cb_{ab}.
\label{H'=H+db}
\end{equation}
It acts also on the tensors $P$, $J$ and $\Phi$ defining an almost
Poisson--quasi--Nijenhuis structure by setting 
\begin{subequations}
\label{btrPJPhi}
\begin{align}
&P'{}^{ab}=P^{ab},
\label{btrPJPhia} 
\\
&J'{}^a{}_b=J^a{}_b-P^{ac}b_{cb},%\vphantom{1\over 2}
\label{btrPJPhib}
\\
&\Phi'{}_{abc}=\Phi_{abc}+\partial_a\phi_{bc}+\partial_b\phi_{ca}+\partial_c\phi_{ab},
\label{btrPJPhic}
\end{align}
where $\phi_{ab}$ is given by  \hphantom{xxxxxxxxxxxxxxxxxxxxxxxxxx}
\begin{equation}
\phi_{ab}=b_{ac}J^c{}_b-b_{bc}J^c{}_a+P^{cd}b_{ca}b_{db}.
\end{equation}
\end{subequations}

It is immediate to see that the BV odd symplectic form
$\Omega_{M,H}$ given in \eqref{OmegaMH} is not invariant
under $b$ transformation \cite {Zucchini1}. To render it invariant, it is
necessary to make $b$ transformation act also on the sigma model fields as
\begin{subequations}
\begin{align}
x'^a&=x^a,
\label{}
\\ 
y'{}_a&=y_a+b_{ab}(x)dx^b.
\label{}
\end{align}
\end{subequations}
One then has \hphantom{xxxxxxxxxxxxxxxxxxxxxxxxxx}
\begin{equation} 
\Omega'{}_{M,H}=\Omega_{M,H}, 
\end{equation}
as required.
It is straightforward to verify that the Hitchin action functionals
$S_{Hr}$ are also both invariant under $b$ transformation,
\begin{equation} 
S'{}_{Hr}=S_{Hr},\qquad r=1,~2.
\end{equation}
This shows that $b$ transformation is a duality symmetry of the Hitchin
model \cite{Zucchini1}. 

$b$ transformation can be rendered a symmetry of the Hitchin--Weil model if we
stipulate further that the tensors $u_i$, $\tau_i$ and $\mu_i$ transform as 
\begin{subequations}
\label{btrutaumu}
\begin{align}
&u'{}_i{}^a=u_i{}^a,
\label{btrutaumua} 
\\
&\tau'{}_{ia}=\tau_{ia}+b_{ab}u_i{}^b,
\label{btrutaumub}
\\
&\mu'{}_i=\mu_i.
\label{btrutaumuc}
\end{align}
\end{subequations}
Upon doing this, it is readily seen that the Hitchin--Weil action functionals
$S_{HWr}$ are also both invariant under $b$ transformation,
\begin{equation} 
S'{}_{HWr}=S_{HWr},\qquad r=1,~2.
\end{equation}

As we shall see, $b$ symmetry plays an important role also in the analysis
of reduction given in the next section. 

%\vfill\eject

\begin{small}
\section{  \bf Geometrical interpretation}
\label{sec:geometry}
\end{small}

Let $M$ be a manifold. An almost Poisson structure on $M$ is
an element $P\in C^\infty(M,$ $\wedge^2TM)$. An  almost Poisson structure $P$ 
is a Poisson structure if 
% satisfying \hphantom{xxxxxxxxxxxx}
\begin{equation}
[P,P]=0,
\label{PP=0}
\end{equation}
where $[~,~]$ denotes the Schoutens--Nijenhius brackets. (More explicitly, 
$[P,P]\in C^\infty(M,\wedge^3TM)$ is given by the right hand side 
of \eqref{AH/gi} below). 
\eqref{PP=0} is nothing but \eqref{Aabc=0} expressed in coordinate free form.
As is well known, when a Poisson structure $P$ on $M$ is given, one can 
define Poisson brackets on $C^\infty(M)$ in standard fashion. 

Assume now that the our Poisson manifold $(M,P)$ carries the action of 
a connected Lie group $G$ with Lie algebra $\mathfrak{g}$ represented 
infinitesimally by the $\mathfrak{g}^\vee$--valued vector field  
$u\in C^\infty(M,\mathfrak{g}^\vee\otimes TM)$. 
The action is said Hamiltonian, if there exist a $\mathfrak{g}^\vee$--valued 
scalar $\mu\in C^\infty(M,\mathfrak{g}^\vee)$, called the moment map,  
such that 
\footnote{$\vphantom{\bigg[}$ Here and below, we view $P$ equivalently 
as a section of $\mathrm{Hom}(T^*M,TM)$.} %\hphantom{xxxxxxxxxxxxxxx}
\begin{subequations}
\label{Predconds}
\begin{align}
&u_i=-Pd_M\mu_i,
\label{Predcondsa}
\\
&l_{u_i}\mu_j=f^k{}_{ij}\mu_k.
\label{Predcondsc}
\end{align}
\end{subequations}
These are precisely conditions \eqref{PWmastercnds} written in intrinsic notation.
When \eqref{PP=0}, \eqref{Predconds} hold, one has \hphantom{xxxxxxxxxxxx}
\begin{subequations}
\label{Predfacts}
\begin{align}
&l_{u_i}P=0,
\label{Predfactsa}
\\
&l_{u_i}u_j-f^k{}_{ij}u_k=0,
\label{Predfactsd}
\end{align}
\end{subequations}
so that $P$ is invariant and the $u$ is equivariant. 
These are relations \eqref{PWrels} upon taking \eqref{Aabc=0},
\eqref{PWmastercnds} into account written again in intrinsic notation.

A classic result of Marsden and Ratiu \cite{Ratiu1}
(see also \cite{Ratiu2}) ensures that, under these conditions, 
if $a\in\mathfrak{g}^\vee$
with coadjoint orbit $\mathcal{O}_a$ and $\mu^{-1}(\mathcal{O}_a)$
is a submanifold of $M$ on which $G$ acts freely
and properly, then the quotient $M_a=\mu^{-1}(\mathcal{O}_a)/G$ inherits a 
Poisson structure $P_a$. Thus, the Poisson--Weil model described 
in sect. \ref{sec:Poisson} encodes Poisson reduction. 

Next, we want to analyze the extent to which the above standard Poisson
reduction framework extends to Poisson--quasi--Nijenhuis structures.
To the best of our knowledge, no such reduction scheme has been been
developed so far. However, since, as shown above, Poisson reduction is encoded 
in the Poisson--Weil model, it is reasonable to expect that  
Poisson--quasi--Nijenhuis reduction may be encoded in the 
Hitchin--Weil model expounded in sect. \ref{sec:Hitchin}. 

Poisson--quasi--Nijenhuis structures were first introduced by Sti\'enon and Xu
in \cite{Xu2}, who, in turn, were inspired by earlier work by Magri e Morosi 
\cite{Magri1}. The authors of \cite{Xu2} considered only the untwisted case, 
but their analysis can be extended to the twisted case directly.

A manifold $M$ is called twisted if it is equipped with a closed 
$3$--form $H\in C^\infty(M,\wedge^3T^*M)$ \hphantom{xxxxxxxxxxxx}
\begin{equation}
d_MH=0.
\label{dH=0/gi}
\end{equation}
Henceforth, we assume that $M$ is twisted. 

An almost Poisson--quasi--Nijenhuis structure on $M$ is a triple
$(J,P,\Phi)$, where $P\in C^\infty(M,\wedge^2TM)$, $J\in C^\infty(M,\mathrm{End}\,TM)$, 
$\Phi\in C^\infty(M,\wedge^3T^*M)$ with 
\begin{equation}
d_M\Phi=0,
\label{dPhi=0/gi}
\end{equation}
(cf. eq. \eqref{dPhi=0}) and satisfying the compatibility condition 
\begin{equation}
JP-PJ^t=0
\label{JPcomp/gi}
\end{equation}
(cf. eq. \eqref{JPcomp}).
An almost Poisson--quasi--Nijenhuis structure $(J,P,\Phi)$ on $M$ is an $H$ twisted 
Poisson--quasi--Nijenhuis structure if 
\begin{subequations}
\label{PqNconds/gi}
\begin{align}
A_H&=0,
\label{AHabc=0/gi}
\\
B_H&=0,
\label{BHabc=0/gi}
\\
C_H&=0,
\label{CHabc=0/gi}
\end{align}
\end{subequations}
where the tensor $A_H\in C^\infty(M, \wedge^3TM)$,
$B_H\in C^\infty(M,\wedge^2TM\otimes T^*M)$,
$C_H\in C^\infty(M,TM\otimes \wedge^2T^*M)$
are defined by
\begin{subequations}
\label{AhBHCH/gi}
\begin{align}
A_H(\alpha,\beta)&=[P\alpha,P\beta]-P\{\alpha,\beta\}_P,
\label{AH/gi}
\\
B_H(\alpha,\beta)&=\{\alpha,J^t\beta\}_P-\{\beta,J^t\alpha\}_P
-\{\alpha,\beta\}_{PJ^t}
-J^t\{\alpha,\beta\}_P+i_{P\alpha}i_{P\beta}H,
\label{BH/gi}
\\
C_H(X,Y)&=[JX,JY]-J\big([JX,Y]-[JY,X]-J[X,Y]\big)
\label{CH/gi}
\\
&\hskip5cm-P\big(i_Xi_Y\Phi-i_{JX}i_YH+i_{JY}i_XH\big),
\nonumber
\end{align}
where $\alpha,\beta\in C^\infty(M,T^*M)$ and $X,Y\in C^\infty(M,TM)$, 
\begin{equation}
\{\alpha,\beta\}_K=l_{K\alpha}\beta
-l_{K\beta}\alpha-\frac{1}{2}d_M(i_{K\alpha}\beta-i_{K\beta}\alpha),
\end{equation}
\end{subequations}
for $K\in C^\infty(M,\wedge^2TM)$, 
and $l$ and $i$ denote Lie derivation and contraction, respectively.
It is straightforward to check that the local coordinate expressions
of $A_H$, $B_H$, $C_H$ are precisely those given by eq. \eqref{AHBHCH},
justifying the claim previously made about the underlying geometry of the
Hitchin model.

In \cite{Xu2}, a further condition is added (in the $H=0$ case).
The $3$--form $\Phi$ is required to satisfy the condition
\begin{equation}
\label{dJPhi=0}
d_J\Phi=0,\qquad (H=0), 
\end{equation}
where $d_J=[J^t\wedge, d]$. To understand the reason of this condition, we
recall the following result proven in \cite{Xu2}. The conditions 
\eqref{PqNconds/gi} together are equivalent to:  $1)$ $(T^*M,\{,\},P)$
being a Lie algebroid; $2)$ $d_J$ being a degree $1$ derivation of the 
associated Gerstenhaber algebra $(C^\infty(M,\wedge^* T^*M),\wedge,[.,.])$;
$3)$ the relation $d_J{}^2=[\Phi,.]$. These three properties
together with \eqref{dJPhi=0} render $(T^*M,\{,\},P, d_J,\Phi)$ a quasi Lie 
bialgebroid. Thus, an untwisted Poisson--quasi Nijenhuis structure on $M$, in
the more restricted sense used here,  
is tantamount of a quasi Lie bialgebroid structure on $T^*M$. The condition 
\eqref{dJPhi=0} is added, among other things, because the
relation $d_J{}^2=[\Phi,.]$ requires as a consistency condition that $[d_J\Phi,.]=0$
and \eqref{dJPhi=0} is sufficient for this to hold. This indicates that the 
three conditions \eqref{PqNconds/gi} imply \eqref{dJPhi=0} or some mild 
generalization of it. As we have seen, \eqref{dJPhi=0} does not
follow from our BV analysis. The classical BV master equation yields 
the conditions which the target space geometry must satisfy for the 
welldefinedness of the model, but of course it does not yield the consistency conditions 
which these imply. 

Poisson--quasi--Nijenhuis geometry is covariant not only under diffeomorphism
symmetry but also under $b$ transformation symmetry. 
For $b\in C^\infty(M,\wedge^2 T^* M)$, the $b$--transform of the 
$3$--form $H$ is 
\begin{equation}
H'=H+d_Mb,
\label{H'=H+db/gi}
\end{equation}
(cf. eq. \eqref{H'=H+db}).
The $b$ transform of an almost Poisson--quasi--Nijenhuis structure $(P,J,\Phi)$ is
given by \hphantom{xxxxxxxxxxxxxxx}
\begin{subequations}
\label{btrPJPhi/gi}
\begin{align}
P'&=P,
\label{btrPJPhi/gia}
\\
J'&=J-Pb,
\label{btrPJPhi/gib}
\\
\Phi'&=\Phi+d_M(J^t\wedge b-bPb),
\label{btrPJPhi/gic}
\end{align}
\end{subequations}
(cf. eq. \eqref{btrPJPhi}). 
It is straightforward though lengthy to verify that $(P,J,\Phi)$ is an 
$H$ twisted Poisson--quasi--Nijenhuis structure, then $(P',J',\Phi')$ is
$H'$ twisted Poisson--quasi--Nijenhuis structure. 

Assume now that the our Poisson--quasi--Nijenhuis manifold $(M,P,J,\Phi)$ 
carries the action of a connected Lie group $G$ with Lie algebra
$\mathfrak{g}$. Intuitively, since the relevant vector bundle in 
Poisson--quasi--Nijenhuis is $TM\oplus T^*M$ rather than $TM$, 
as in generalized complex geometry, we expect that  
the $G$ action is represented at the infinitesimal level not only by a 
$\mathfrak{g}^\vee$--valued vector field $u\in C^\infty(M,\mathfrak{g}^\vee\otimes TM)$, 
as above, but also by a $\mathfrak{g}^\vee$--valued $1$--form 
$\tau\in C^\infty(M,\mathfrak{g}^\vee\otimes T^*M)$, which we name moment
$1$--form in compliance with common usage. 
We call the $G$ action Hamiltonian, if there exist a
$\mathfrak{g}^\vee$--valued scalar 
$\mu\in C^\infty(M,\mathfrak{g}^\vee)$, called the moment map,  
such that 
\begin{subequations}
\label{redconds}
\begin{align}
&u_i=-Pd_M\mu_i,
\label{redcondsa}
\\
&\tau_i=J^td_M\mu_i,
\label{redcondsb}
\\
&l_{u_i}\mu_j=f^k{}_{ij}\mu_k.
\label{redcondsc}
\end{align}
\end{subequations}
These are precisely conditions \eqref{HWmastercnds} written in intrinsic
notation. They generalize \eqref{Predconds} in obvious fashion. 
When \eqref{redconds}, \eqref{PqNconds/gi} hold, 
\begin{subequations}
\label{redfacts}
\begin{align}
&l_{u_i}P=0,
\label{redfactsa}
\\
&l_{u_i}J-P\Upsilon_i=0
\label{redfactsb}
\\
&i_{u_i}\Phi-d_M\Xi_i+J^t\wedge\Upsilon_i=0
\label{redfactsc}
\\
&l_{u_i}u_j-f^k{}_{ij}u_k=0,
\label{redfactsd}
%\\
\end{align}
\begin{align}
&l_{u_i}\tau_j-f^k{}_{ij}\tau_k-i_{u_j}\Upsilon_i=0,
\label{redfactsg}
\end{align}
where $\Xi\in C^\infty(M,\mathfrak{g}^\vee\otimes T^*M)$, 
$\Upsilon\in C^\infty(M,\mathfrak{g}^\vee\otimes \wedge^2T^*M)$ are given by
\begin{align}
\Xi_i&=(1+J^tJ^t)d_M\mu_i,
\label{redfactse}
\\
\Upsilon_i&=d_M\tau_i-i_{u_i}H.
\label{redfactsf}
\end{align}
\end{subequations}
These are relations \eqref{HWrels} upon taking \eqref{PqNconds}, 
\eqref{HWmastercnds} into account written again in intrinsic notation.
They generalize \eqref{Predfacts} in a rather non trivial way.
We see that $H$ is not invariant and that, while $P$ is 
invariant, $J$, $\Phi$ fail to be so. Similarly, while  $u$ is equivariant, 
$\tau$ is not. In all cases, the obstruction is given by the $2$--form
$\Upsilon$.

In the presence of a $G$ action on $M$, the above geometric framework is
covariant under $b$ transformation provided this acts also on $u$, $\tau$ and
$\mu$ as 
\begin{subequations}
\label{btrutaumu/gi}
\begin{align}
u'_i&=u_i,
\label{btrutaumu/gia}
\\
\tau'_i&=\tau_i-i_{u_i}b,
\label{btrutaumu/gib}
\\
\mu'_i&=\mu_i,
\label{btrutaumu/gic}
\end{align}
\end{subequations}
(cf. eq. \eqref{btrutaumu}).
From these relations and from \eqref{redfacts}, one realizes immediately 
that the failure of $H$, $J$, $\Phi$ to be invariant and, similarly, of $\tau$ 
to be equivariant has the form of an infinitesimal $b$ transform with 
$b=\Upsilon_i$ for given $i$. That this comes about is hardly surprising, given the 
$b$ symmetry of the Hitchin--Weil model, from which \eqref{redfacts}
were obtained. It reflects also the fact that the symmetry of 
the geometry considered here is larger than the diffeomorphism
one and contains also $b$ transformation, as in generalized complex geometry. 
The natural question arises about
whether it is possible to make all the $\Upsilon_i$ vanish by means of a 
single $b$ transform. It is easy to see that, to this end,  it is sufficient 
that the $b$ field solves the equation \hphantom{xxxxxxxxxxxxxxxxxxxxxxxx}
\begin{equation}
l_{u_i}b=\Upsilon_i.
\label{lub=Ups}
\end{equation}
Unfortunately, general conditions under which \eqref{lub=Ups} 
has solutions are not known to us. Alternatively, one may impose the condition
\begin{equation}
\Upsilon_i=0,
\label{Ups=0}
\end{equation}
by hand. This, however, is not yielded by the formalism in natural fashion.
It is natural to wonder whether the above provides a viable
framework for the reduction of Poisson--quasi--Nijenhuis structures.
We have no answer as yet. It is however useful to that end 
to examine what is known about reduction in generalized complex geometry.

A generalized almost complex structure $\mathcal{J}$ is a section of 
$C^\infty(\mathrm{End}(TM\oplus T^* M))$, which is an isometry of 
the natural Courant metric $\langle~ ,~\rangle$ of $TM\oplus T^* M$ and 
satisfies \hphantom{xxxxxxxxxxxxxxxxxxxxxxxxxxxxxxx}
\begin{equation}
\mathcal{J}^2=-1
\label{JJ2=-1}
\end{equation}
\cite{Gualtieri}.
The generalized almost complex structure $\mathcal{J}$ is called a 
generalized complex structure if its $+i$ eigenbundles $L_\mathcal{J}$ of
$\mathcal{J}$ is involutive with respect to the $H$ twisted Courant brackets $[\,,]_H$
of $TM\oplus T^* M$ \cite{Gualtieri}
\footnote{$\vphantom{\bigg[}$ The $\pm i$ eigenbundles of $\cal J$ are complex and, thus, 
their analysis requires complexifying $TM\oplus T^* M$ leading to 
$(TM\oplus T^* M)\otimes\mathbb{C}$.}.

It is often convenient to write a generalized almost complex structure
$\mathcal{J}$ in the block form \hphantom{xxxxxxxxxxxxxxxxxxxxxxxxxxxxxxxxx}
\begin{equation}\label{calJ}
\mathcal{J}=\bigg(\begin{matrix}J& Q\\ P& -J^t\end{matrix}\bigg),
\end{equation}
where $P\in C^\infty(M,\wedge^2 TM)$, $J\in C^\infty(M,\mathrm{End}TM)$, 
$Q\in C^\infty(M,\wedge^2 T^* M)$. It is easily checked that
the triple $(P,J,\Phi)$, where
\begin{equation}\Phi=d_MQ,
\label{Phi=dMQ/gi}
\end{equation}
is an almost Poisson--quasi--Nijenhuis structure satisfying besides \eqref{JPcomp/gi}
two more algebraic conditions following from \eqref{JJ2=-1} and corresponding
to eq. \eqref{JPQcomp}. If $\mathcal{J}$ is a generalized complex
structure, then $(P,J,\Phi)$ is a Poisson--quasi--Nijenhuis structure
satisfying besides \eqref{PqNconds/gi} an extra differential condition 
following from Courant involutivity of $L_\mathcal{J}$ 
and corresponding to eq. \eqref{DH=0}.

Assume now that our generalized complex manifold $(M,\mathcal{J})$ carries the 
the action of a connected Lie group $G$ with Lie algebra $\mathfrak{g}$
represented infinitesimally by the vector fields $u_i$.
Following Lin and Tolman \cite{Tolman1,Tolman2}
(see also \cite{Hu}), we define a generalized moment map to be an element 
$A\in C^\infty(M,\mathfrak{g}^\vee\otimes(TM\oplus T^* M)\otimes\mathbb{C})$ 
of the form \hphantom{xxxxxxxxxxxxxxxx}
\begin{equation}
A_i=u_i+\tau_i-id_M\mu_i 
\end{equation}
such that \hphantom{xxxxxxxxxxxxxxxx}
\begin{equation}
\mathcal{J}A_i=iA_i
\label{JJa=iA}
\end{equation}
and that \eqref{redcondsc} holds. It is easy to see that
\eqref{JJa=iA} implies \eqref{redcondsa}, \eqref{redcondsb} and summarizes 
in intrinsic form \eqref{HWrelsf}, \eqref{Wi=0}. 

Let us assume that \eqref{Ups=0} holds. \eqref{Ups=0} is just \eqref{LTcnds}.
From \eqref{redfactsa}--\eqref{redfactsc} and \eqref{Ups=0}, 
it follows that $H$ and $P$, $J$, $Q$ and, so, $\mathcal{J}$ are all invariant. 
Similarly, \eqref{redfactsg} and \eqref{Ups=0} imply that $\tau$ 
is equivariant. According to the authors of \cite{Tolman1,Tolman2},
under these conditions, if, for $a\in\mathfrak{g}^\vee$ with coadjoint orbit 
$\mathcal{O}_a$ and if $\mu^{-1}(\mathcal{O}_a)$ is a submanifold of $M$ 
on which $G$ acts freely, then the quotient $M_a=\mu^{-1}(\mathcal{O}_a)/G$ 
inherits a generalized complex structure $\mathcal{J}_a$. 

The above analysis shows that the reduction scheme of Lin and Tolman 
is a particular case of the one worked out in this paper. It  
seems therefore to point to a reduction framework 
far more general than that considered by Lin and Tolman. One one hand, it may apply to
Poisson--quasi--Nijenhuis structures, which are more general than
generalized complex ones. On the other, strict invariance may not be necessary
at the end and the weaker conditions \eqref{redfactsa}--\eqref{redfactsc} 
may suffice.

%\vfill\eject

\begin{small}
\section{  \bf Discussion}
\label{sec:remarks}
\end{small}

In sects. \ref{sec:Poisson}, \ref{sec:Hitchin}, we have argued that
the Poisson--Weil and Hitchin--Weil sigma models encode the symmetry 
reduction of the Poisson and Hitchin sigma models, respectively.
In a sense, coupling to the Weil model should perform the same type 
of function as gauging and may be considered to be a gauging in a sense, 
though, strictly speaking, there is no gauge field that interacts with the 
ungauged sigma model fields. 

The sigma models studied in this paper cannot be considered
fully fledged quantum field theories as long as gauge fixing
is not carried out, since, in the absence of gauge fixing, the kinetic terms
of the fields are ill defined. Fixing the gauge requires restricting the fields
on a suitable functional submanifold $\mathfrak{L}$ in field space, 
that is Lagrangian with respect to the BV odd symplectic form
\cite{BV1,BV2,AKSZ}. The restriction results in certain relations among 
the fields. Formal arguments, based on the BV master equation, 
indicate that the resulting gauge fixed  
field theory is independent at the quantum level 
from the choice of $\mathfrak{L}$ as long as 
the choices considered can be continuously deformed 
one into another. Unfortunately, fixing the gauge is usually a technically very 
hard problem \cite{AKSZ,Zucchini3}.

We have seen that symmetry reduction of a Poisson or a generalized complex
manifold requires the choice of some element $a\in\mathfrak{g}^\vee$. 
The reduced manifold is then
the quotient $M_a=\mu^{-1}(\mathcal{O}_a)/G$, where $\mathcal{O}_a$ 
is the coadjoint orbit of $a$. However, there is no trace of such
a choice in the models we described. It is likely that $a$ enters 
in some way in the definition of the functional Lagrangian submanifold $\mathfrak{L}$ 
involved in gauge fixing. However, at the moment, this is only a speculation.
Clearly, much work remains to be done to reach a better 
understanding of these matters.

\vskip .5cm

{\bf Acknowledgments. } We thank J. Louis for organizing thr Workshop on
"Generalized Geometry and Flux Compactifications" 
held in DESY from February 19th through March 1st 2007, 
during which this paper was first conceived. 
We thank J.-P. Ortega for correspondence 
and useful suggestions.

%\vskip .5cm

\vfill\eject

\appendix

\begin{small}
\section{  \bf De Rham superfields}
\label{sec:superfields}
\end{small}

In general, the fields of a 2--dimensional field theory are differential 
forms on a oriented closed $2$--dimensional manifold $\Sigma$. They can be viewed 
as elements of the space $C^\infty(T[1]\Sigma)$ of functions on the 
Grassmann degree $1$ 
tangent bundle $T[1]\Sigma$ of $\Sigma$, which we shall call 
de Rham superfields. 
More explicitly, we associate with the coordinates $z^\alpha$ of 
$\Sigma$ Grassmann odd partners $\zeta^\alpha$ with 
\begin{equation}
\deg z^\alpha=0, \qquad \deg\zeta^\alpha=1.\vphantom{\Big[}
\label{}
\end{equation}
$T[1]\Sigma$ is endowed with a natural differential $d$ defined by 
\begin{equation}
dz^\alpha=\zeta^\alpha,\qquad d\zeta^\alpha=0.\vphantom{\Big[}
\label{}
\end{equation}
A generic de Rham superfield $\psi(z,\zeta)$ is a triplet 
formed by a $0$--, $1$--, $2$--form field $\psi^{(0)}(z)$, 
$\psi^{(1)}{}_\alpha(z)$, $\psi^{(2)}{}_{\alpha\beta}(z)$ 
organized as
\begin{equation}
\psi(z,\zeta)=\psi^{(0)}(z)+\zeta^\alpha\psi^{(1)}{}_\alpha(z)
+\frac{1}{2}\zeta^\alpha\zeta^\beta\psi^{(2)}{}_{\alpha\beta}(z).
\label{}
\end{equation}
The forms $\psi^{(0)}$, $\psi^{(1)}$, $\psi^{(2)}$ are called the components 
of $\psi$. Note that, in this formalism, the exterior differential of $\Sigma$ 
can be identified with the operator 
\begin{equation}
d=\zeta^\alpha\partial/\partial z^\alpha.
\label{}
\end{equation}

The coordinate invariant integration measure of $T[1]\Sigma$ is 
\begin{equation}
\varrho=dz^1dz^2d\zeta^1d\zeta^2.
\label{measure}
\end{equation}
Any de Rham superfield $\psi$ can be integrated on $T[1]\Sigma$ according to
the prescription \hphantom{xxxxxxxxxxxxxxxxxxxxxxx}
\begin{equation}
\int_{T[1]\Sigma}\varrho\,\psi=\int_\Sigma\frac{1}{2}\,
dz^\alpha dz^\beta\psi^{(2)}{}_{\alpha\beta}(z).
\label{}
\end{equation}
By Stokes' theorem,  \hphantom{xxxxxxxxxxxxxxxxxxxxxxx}
\begin{equation}
\int_{T[1]\Sigma}\varrho\, d\psi=0.
\label{}
\end{equation}

It is possible to define functional derivatives of functionals of de Rham superfields.
Let $\psi$ be a de Rham superfield and let $F(\psi)$ be a functional of $\psi$.
We define the left/right functional derivative superfields 
$\delta_{l,r} F(\psi)/\delta\psi$ as follows. 
Let $\sigma$ be a superfield of the same properties as $\psi$.
Then, 
\begin{equation}
\frac{d}{dt}F(\psi+t\sigma)\Big|_{t=0}
=\int_{T[1]\Sigma}\varrho\,\sigma\frac{\delta_l F(\psi)}{\delta\psi}
=\int_{T[1]\Sigma}\varrho\,\frac{\delta_r F(\psi)}{\delta\psi}\sigma.
\label{}
\end{equation}

In the applications below, the components of the relevant de Rham superfields
carry, besides the form degree, also a ghost degree. We shall limit ourselves 
to homogeneous superfields. A de Rham superfield $\psi$ is said homogeneous 
if the sum of the form and ghost degree is the same for all its components 
$\psi^{(0)}$, $\psi^{(1)}$, $\psi^{(2)}$ of $\psi$. 
The common value of that sum is called the (total) degree $\deg\psi$ of $\psi$. 
It is easy to see that the differential operator $d$ and the
integration operator $\int_{T[1]\Sigma}\varrho$ carry 
degree $1$ and $-2$, respectively. 
Also, if $F(\psi)$ is a functional of a superfield $\psi$, then
$\deg \delta_{l,r} F(\psi)/\delta\psi=\deg F-\deg \psi+2$.

%\vfill\eject

\begin{small}
\section{  \bf The functional derivation $\delta/\delta x^a$}
\label{sec:techno}
\end{small}
Since, for given $x\in C^\infty(T[1]\Sigma, M)$,  
one has $y\in C^\infty(T[1]\Sigma, x^*T^*[1]M)$, it is not possible to vary 
$x$ keeping $y$ fixed. In fact, the condition $\delta y=0$ is not covariant, 
as is easy to see, and, so, it cannot be consistently imposed. 
This poses a technical problem for the computation of the functional
derivatives $\delta F/\delta x^a$, when $F$ explicitly depends on $y$.
The difficulty is solved by picking a connection $\Gamma$ of $M$ 
and requiring that 
\begin{equation}
\delta_\mathrm{cov}y_a=\delta y_a-\Gamma^b{}_{ca}(x)\delta x^cy_b=0,
\label{deltacovy=0}
\end{equation}
under variation of $x$. It is convenient to take $\Gamma$ torsionless.
One then computes $\delta_\mathrm{cov}F/\delta x^a$ by varying both $x$ and
$y$ with $\delta y$ given by \eqref{deltacovy=0}. 
The result depends of course on the choice $\Gamma$. 
However, in all the relevant calculations, $\Gamma$ drops out at the end,  
reflecting the intrinsic covariance of the theory. 

%\vfill\eject

The BV brackets \eqref{Mantbrckt}, \eqref{MHantbrckt} are to be computed 
by replacing $\delta/\delta x^a$ by $\delta_\mathrm{cov}/\delta x^a$ throughout.
It can be checked that the result does not depend on $\Gamma$. 
Similarly, if $S_t$ is a BV master action, then the BV variations, obtained from 
\begin{subequations}
\begin{align}
&\delta_t x^a=(S_t,x^a),
\label{}
\\
&\delta_t y_a-\Gamma^b{}_{ca}(x)\delta_tx^cy_b=(S_t,y_a),
\label{}
\end{align}
\end{subequations}
also do not depend on $\Gamma$.

%\vfill\eject

\end{document}